\documentclass{article}
\usepackage{graphicx}
\newcommand{\exv}[1]{\left\langle{#1}\right\rangle}

\begin{document}

\title{Quantum corrections to Boltzmann equations}
\author{A. Jakov\'ac\footnote{Antal.Jakovac@cern.ch}\\
Institute of Physics, Technical University of Budapest,
    Budafoki \'ut. 8\\ H-1111 Budapest, Hungary}

\maketitle

\section*{Abstract}

Because of IR (pinch) singularities a resummation is
necessary in non-equilibrium field theories, that can be performed
by using Kadanoff--Baym equations. Taking Landau prescription
correctly into account, Kadanoff--Baym equations reduce to Boltzmann
equations only in a restricted kinematical range; in other cases a
new equation (the former constraint equation) has to be
considered. In relaxation time approximation this new equation
results in the shifting and smearing of multiparticle thresholds.

\section{Introduction}

The application of Boltzmann equations is a generally accepted tool
for studying non-equilibrium processes in different domains of
physics. In field theory one can justify this method by using
Kadanoff--Baym equations\cite{kadbaym}. This method presents,
however, just an indication that Boltzmann equation is consistent with
field theory, but neither embeds it into field theory, nor proves that
there cannot be other important contributions from different sources.

If we try to use Boltzmann equations consistently with perturbation
theory, we have to investigate which is that subset of the Feynman
series that is responsible for the appearance of Boltzmann equations,
why is it relevant, and what is the effect of the terms left out from
this subset. It is known for long times that in statistical physics
one can obtain results from diagrammatic approach which is consistent
with Boltzmann equations if we resum ladder diagrams\cite{Mahan}.
Recently it has been demonstrated that ladder and extended ladder
diagrams give important contribution to composite operator expectation
values in scalar theories, and the results are consistent with
Boltzmann equations\cite{ladder}.

What is that effect that renders the ladder diagrams so important? As
it was proved in\cite{J1}, each pairs of propagators in a ladder
yields pinch singular contributions\cite{pinch}. In real space these
diagrams give secular terms (proportional to some power of time);
after resummation they result in Boltzmann equations\cite{BVW}.

Boltzmann equations in linear response theory yield exponential
damping. Still, in several numerical studies one observes power law
damping of different physical quantities\cite{powerlaw}. In some
systems these are consequences of non-linear, hydrodynamic effects
\cite{Yaff}. However, many particle (quantum) coherence can give such
terms even in linear response regime\cite{manypart}. These effects
clearly cannot be part of the Boltzmann solution.

In this contribution we try to show that, from field theoretical point
of view, linearized Kadanoff--Baym equations represent a tool for
resummation the pinch singular parts of the ladder diagrams. We will
argue that for long time evolution of a quasi-conserved quantity
exactly these are the diagrams that are relevant for the exponential
damping, and that the solution of Boltzmann equations yields correct
damping rate. We will argue furthermore that, besides the Boltzmann
contribution, there can be other effects coming from multiparticle
thresholds that yield exponentially damped power law time dependence.
For more details see\cite{J1,J2}.

\section{Long time behavior}

First we produce the out of equilibrium state from where the
non-equilibrium time dependence can be started. We can do it by
starting from equilibrium and modifying the time evolution for a
certain time as $H\to H + \Delta H$. We denote $\Delta H=\int \hat F
\zeta(x)$, where $\hat F$ is a local operator, $\zeta(x)$ denotes its
strength. Assuming small deviation from equilibrium we can use linear
response in $Delta H$, i.e. for an arbitrary operator having zero
expectation value in equilibrium we can write
\begin{equation}
  \exv{O(x)} = \int d^4x' G_R^{OF}(x-x') \zeta(x'),
\end{equation}
where $G_R^{OF}(x) = i\,\Theta(t) \exv{[O(x),\hat F(x')]}_{eq}$. It is
similar to the Kubo formula; however, if we fix only finite number of
initial observables, we have a big freedom in the choice of $\Delta
H$. The best we can expect is that after long times we observe
universal time dependence.

Under some assumptions\cite{J1} we can rewrite these expressions as
(suppressing the spatial dependence)
\begin{equation}
  \exv{O(t)} = \int\frac{dk_0}{2i\pi} e^{-ik_0 t} \rho^{OF}(k_0)
  \zeta(k_0),
\end{equation}
where $\rho^{OF}(k_0)=\mathrm{Disc} i G_R^{OF}(k_0)$. Analyzing this
expression for long times we obtain two possible contributions: one
from saddle points (maxima), the other from thresholds. The former
yields exponential, the latter yields power law time dependence.

If $O$ is conserved at tree level then
$\rho_0^{OF}(k_0)\sim\delta(k_0)$. Interactions spoil the
conservation, but, if it is quasi-conserved, the trace of the tree
level peak should remain. Then we expect the form
\begin{equation}
   \exv{O(t)} = Q_1 e^{-\Gamma t} + Q_2 t^{-\alpha+1} \cos(\Omega
   t+\phi).
\end{equation}
Here $Q_1,\,Q_2$ and $\phi$ depend on the choice of $\Delta H$, ie. on
initial conditions; $\Gamma$ (the width of the peak at zero momentum),
$\alpha$ and $\Omega$ (the index and position of the threshold),
however, we expect to be universal quantities.

\section{Pinch singularities and resummation}

The appearance of the damping means IR divergences for an infinite
subset of diagrams. This is because a broad peak at zero momentum
should come from a Green's function that behaves as $G_R^{OF}(k_0)\sim
(k_0+i\Gamma)^{-1}$ near the peak. Since $\Gamma\sim g^\#$ where $g$
is the coupling, this behavior must show up in perturbation theory as
$(1/k_0) \sum_n (-i\Gamma/k_0)^n$. Therefore in $n$th order of
perturbation theory we expect $k_0^{-n-1}$ pole for small momenta.

To identify the origin of this IR behavior in the perturbative
analysis we consider ladder diagrams in R/A formalism. Multiplying a
retarded and an advanced propagator we find singular behavior if the
external momentum vanishes; up to non-singular terms
\begin{equation}
  G_R(Q-\frac K2)G_A(Q+\frac K2) \to i\rho(Q-\frac K2)\left(
  \frac{\Theta(|2q_0|\!-\!|k_0|)}{QK} +
  \frac{\Theta(|k_0|\!-\!|2q_0|)}{Q^2+\frac{K^2}4-m^2}\right),
\end{equation}
applying Landau prescription. Both equations may produce IR
divergence, when $K=0$ or when $Q^2+\frac{K^2}4-m^2=0$. In a ladder
diagram with $n$ rungs this divergence appears on the $n$th power.
Thus for $\mathbf{k} =0$ we find the same divergence structure as we
expected from the finite width.

\begin{figure}[htbp]
  \centering
  \includegraphics[height=6cm,angle=270]{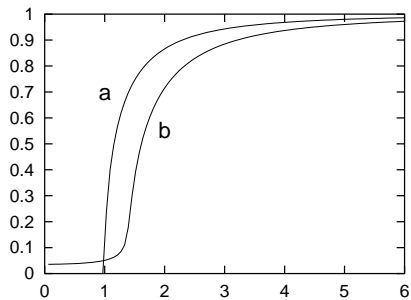}
  \caption{\emph{a)} free threshold; \emph{b)} effect of the
  relaxation time approximation.} 
  \label{fig:dressedthreshold}
\end{figure}

After identifying the IR divergences we can think about their
resummation. Since ladder diagrams are recursive structures we should
find an equation for the non-equilibrium two-point function of the
form
\begin{equation}
  \begin{array}[c]{ll}
    (Q^2+{K^2}/4-m^2) \bar G(Q,K) = ({\emph{A}}_+* \bar G)(Q,K) \qquad
    & \mathrm{for}\quad |k_0/2|>|q_0|\cr 
    2QK \bar G(Q,K) = ({\emph{A}}_-* \bar G)(Q,K) \qquad &
    \mathrm{for}\quad |k_0/2|<|q_0|,\cr
  \end{array}
\end{equation}
where ${\emph{A}}_\pm$ are linear kernels, non-zero in the pinch singular
limit. Exactly this is the form of the Kadanoff--Baym equations, so we
can use them for the resummation procedure. In a very generic form we
can start from the Schwinger--Dyson equations
\begin{equation}
  (p^2\!-\!m^2)G(p,q) = {\emph{F}}_1(G),\quad(q^2\!-\!m^2)G(p,q) =
  {\emph{F}}_2(G),
\end{equation}
perform Wigner transformation ($p,q = Q\pm K/2$) and take their
difference and sum
\begin{equation}
  \begin{array}[c]{ll}
    \textrm{(diff)}\qquad &2QK \bar G = {\emph{F}}_1(G) -
    {\emph{F}}_2(G)\cr \textrm{(sum)}& 2(Q^2+K^2/4-m^2)\bar G =
    {\emph{F}}_1(G) + {\emph{F}}_2(G)).\cr 
  \end{array}
\end{equation}
The first (diff) equation is the Boltzmann equation for $K\to 0$, and
as it could be seen from the previous analysis, it can be applied for
$|k_0/2|<|q_0|$. The second (sum) equation, which is plays the role of
the on-shellness condition in case of the usual Boltzmann equations,
for $|k_0/2|>|q_0|$ becomes a differential equation. For the free case
it generates a two-particle cut, therefore it yields a power law time
dependence with $\alpha=1/2$. In the interacting case the solution is
rather complicated, but as a first study we can try to adapt
relaxation time approximation for it, and solve
\begin{equation}
  \left[\frac12(k_0+i\Gamma_c)^2 -{\bf Q}^2 - m_{eff}^2\right] G
  = \mathrm{initial\ conditions}.
\end{equation}
The effect of the solution can be seen on
Fig. \ref{fig:dressedthreshold}. 
The threshold is shifted and smeared out. The non-analyticity
disappears, and the original $t^{-3/2}$ time dependence becomes
$e^{-\Gamma_c t} t^{-1}$. The damping factor $\Gamma_c$ is different
from the Boltzmann damping rate, so this term, in principle, can be
dominating for long times.

\section{Conclusions}

\hspace*{\parindent}\emph{Using Boltzmann equations in QFT}: In
perturbation theory IR (pinch) divergences at finite temperature make
resummation necessary. The relevant diagrams are the ladders, their
pinch singular part can be resummed using Boltzmann and threshold
equations.

\emph{Using QFT to correct Boltzmann equations:} To describe long time
behavior, besides the Boltzmann equation we have to take into account
the other Kadanoff--Baym equation as differential equation.  Both
equations have to be considered in their respective validity range in
the momentum space.

\emph{In relaxation time approximation}, we have a ``dressed''
threshold: shifted and smeared out as compared with the undressed
function. Instead of power law time evolution of the collisionless
approximation we find damped power law time evolution.

\section*{Acknowledgment}
This work was partially supported by the Hungarian Science Fund
(\uppercase{OTKA}).

\end{document}